\documentclass[article]{revtex4}
\usepackage{amssymb,amsmath,calrsfs}
\usepackage{epsfig}
\newcommand{\comment}[1]{} 
\newcommand{\nn}{\nonumber}
\def\ll{\mathcal L}
\comment{
}

\numberwithin{equation}{section}

\begin{document}

\title{New relations between spinor and scalar one-loop effective Lagrangians
in constant background fields}

\author{Adolfo Huet}
\email{ huet@phys.uconn.edu} \affiliation{Department of
Physics\\University of Connecticut\\2152 Hillside Road, U-3046 \\ Storrs, CT 06269, USA}
\date{\today}

\begin{abstract}
Simple new relations are presented between the one-loop effective Lagrangians of spinor
 and scalar particles in constant curvature background fields, both electromagentic and gravitational.
 These relations go beyond the well-known cases for self-dual background fields.
\end{abstract}

\maketitle

\section{Introduction}

The QED one-loop effective Lagrangian was obtained for the first
time by Heisenberg who considered the effect that the Dirac sea
would have on the dynamics of electromagnetic fields, and summarized
this effect as a correction to the classical Maxwell Lagrangian ~\cite{Heisenberg:1935qt}.
The effective Lagrangian corresponding to scalar QED was obtained
shortly afterwards by Weisskopf ~\cite{Weisskopf:1936}.
The effect of  the quantum vacuum, encoded in the effective Lagrangian,
 accounts for physical phenomena  such as pair-production from
vacuum, light-light scattering and vacuum
birefringence, among others~\cite{Dittrich:2000zu}. Using the proper-time
representation \cite{Schwinger:1951nm}, and using Schwinger's choice
of units, we write both  effective Lagrangians  in modern notation~\cite{Dunne:2004nc}:
\begin{eqnarray}
\ll_{{\rm spinor}}( E, B)
=  -\frac{1}{8 \pi^2} \int_0^{\infty} \, \frac{ds}{s^3}\, e^{-m^2 s }
\Big\{ E B s^2\, {\rm cot}(E s)  {\rm coth}(B s)
- 1 - \frac{ s^2}{3}\,( B^2 -E^2 ) \Big\} \, , \label{EHsp}
\end{eqnarray}
\begin{eqnarray}
\ll_{{\rm scalar}}( E, B)
=  \frac{1}{16 \pi^2} \int_0^{\infty} \, \frac{ds}{s^3}\, e^{-m^2 s }
\Bigg\{ \, \frac{ E B\, s^2}{{\rm sin}( E s)  {\rm sinh}( B s)}
- 1 + \frac{ s^2}{6}\,( B^2 -E^2 ) \Bigg\} \label{EHsc}  \,.
\end{eqnarray}
These expressions are the
first (one-loop) quantum corrections  to the classical Maxwell
Lagrangian.
We notice a similar structure in both expressions. Both parametric integrals
contain three types of terms. The main term involves trigonometric
and hyperbolic functions.
We also have the "$-1$" term inside the curly brackets which
corresponds to a subtraction  of the free-field ($E = B = 0$)
Lagrangian and ensures the vanishing of the full expression in the
absence of a background field. And in each case, the last term,
which is proportional to the classical Maxwell Lagrangian,
corresponds to charge renormalization~\cite{Heisenberg:1935qt,Weisskopf:1936,Schwinger:1951nm} .

Looking at both  spinor and scalar Lagrangians,  we notice a
similar structure in terms of certain trigonometric functions. In
fact, simple relations between the spinor and scalar effective Lagrangians occur
for a self-dual background field, or a field of definite helicity~\cite{Duff:1979dy,Dunne:2002qf}.
In Minkowski space this means $E = \pm i B$, and by inspection of  (\ref{EHsp}) and (\ref{EHsc})
we find
\begin{equation} \label{SDrelation1}
\ll_{\rm spinor}(\pm i B, B)=-2\, \ll_{\rm scalar}(\pm i B, B) \, ,
\end{equation}
as follows from the trigonometric identity $\coth^2(x)=1+ 1/{\rm {sinh}}^2(x)$ .

 This relation  reflects  the \emph{isospectrality} of the Dirac and Klein-Gordon
 operators for a self-dual background gauge field~: it is a consequence \cite{Jackiw:1977pu} of the self-duality of the background
that apart from zero-modes, the Dirac operator has exactly the same
spectrum as the corresponding Klein-Gordon operator, with a
multiplicity factor of 4. Since these  one-loop effective Lagrangians can be
expressed  in terms of logarithms of the determinants of
the Dirac and Klein-Gordon operators, this means that, due to the isospectrality,   when the background is self-dual, we have
\begin{equation}  \label{spinorscalar1}
\ll_{\rm spinor}(\pm i B, B)=-2\, \ll_{\rm scalar}(\pm i B, B) + \frac{1}{2}
\Bigg( \frac{e B}{2 \pi} \Bigg)^2 \ln \Bigg( \frac{m^2}{\mu^2}
\Bigg) \, ,
\end{equation}
where $N_0 = ( \frac{e B}{2 \pi} )^2$ is the zero-mode number
density.   When we renormalize on-shell (i.e. $\mu^2 = m^2$ )
we recover (\ref{SDrelation1}). An example of a consequence of this
relation  is that the ${\mathcal N}=2$ SUSY effective Lagrangian
vanishes at one-loop~:
\begin{eqnarray}
\ll_{\rm N=2\; SUSY}(\pm i B, B) &\equiv& \ll_{\rm spinor}(\pm i B, B)+2\, \ll_{\rm scalar}(\pm i B, B) \nn \\
&=& 0 \, .
\end{eqnarray}
This kind of relation has been exploited in order to calculate the effective action, for a fermion
in an  instanton background \cite{'tHooft:1976fv} .

In this paper we present new relations between $ \ll_{\rm spinor}$ and
$\ll_{\rm scalar}$ that apply  when the background is not
self-dual.

\section{Spinor/Scalar relations in QED}

In this section  we use trigonometric
identities to derive spinor/scalar relations for three physically
interesting types of background configurations, including a purely
magnetic background field, a purely electric field (where Schwinger
pair production is possible) and the general case of a constant
electromagnetic field.

\label{QEDrelations}
\subsection{The case of a purely magnetic background}
\label{magnetic} For a constant magnetic field background of
strength $B$, the spinor and scalar effective Lagrangians are :
\begin{eqnarray}
\ll_{{\rm spinor}}(B) &=& - \frac{ B}{8\pi^2}\, \int_{0}^\infty \,
\frac{ds}{s^2} \; {\rm e}^{-m^2 s} \; \left( {\rm coth}\,   B s   -
\frac{1}{ B s} - \frac{ B s}{3} \right)\, ,
\label{1lsp}\\
\ll_{{\rm scalar}}(B)
&=&
 \frac{B}{16\pi^2}\, \int_{0}^\infty \, \frac{ds}{s^2}
\; {\rm e}^{-m^2 s} \; \left( \frac{1}{{\rm sinh}\, B s}   -
\frac{1}{B s} + \frac{B s}{6} \right) \, . \label{1lsc}
\end{eqnarray}
 The simple trigonometric identity
\begin{eqnarray}
\left({\rm coth}\, s   - \frac{1}{ s} - \frac{s}{3}\right)
&=&
\left({\rm coth}\, 2 s   - \frac{1}{2 s} - \frac{2 s}{3}\right)
 +
\left(\frac{1}{{\rm sinh}\, 2 s}   - \frac{1}{2 s} + \frac{2
s}{6}\right)\, , \label{trig1}
\end{eqnarray}
reveals a relation between the spinor and scalar Lagrangians~:
\begin{eqnarray} \label{basicB}
\ll_{{\rm spinor}}(B)
&=&  \frac{1}{2} \ll_{{\rm spinor}}(2 B) -  \ll_{{\rm scalar}}(2 B) \, .
\label{basicB}
\end{eqnarray}
Iterating this relation $N$ times,
we find
\begin{eqnarray} \label{b-iterated}
\ll_{{\rm spinor}}(B)
&=& \frac{1}{2}\left( \, \frac{1}{2} \, \ll_{{\rm spinor}}(4 B) -  \ll_{{\rm scalar}}(4 B)
\, \right) - \ll_{{\rm scalar}}(2 B) \nn \\
&\vdots&  \nn  \\
&=& \frac{1}{2^N} \ll_{{\rm spinor}}(2^N B)
- \sum_{k=1}^N \frac{1}{2^{k-1}} \, \ll_{{\rm scalar}} (2^k B)  \,.
\end{eqnarray}
Rescaling $B$ by $2^{-N}$, we can write this as~:
\begin{eqnarray*}
\ll_{ {\rm spinor} }(B)
 &=&  2^N \ll_{{\rm spinor}}(2^{-N} B )
+ \sum_{l=1}^N  2^l \, \ll_{{\rm scalar}}( 2^{1-l} B)  \, .
\end{eqnarray*}
From the perturbative expansion  of the Lagrangians  it is clear that the leading term in the weak-field limit gives $ \ll(B) \propto B^4$.
Therefore,
 we can take the $N\to\infty$ limit and relate the spinor effective Lagrangian
 to an infinite sum of scalar effective Lagrangians~:
\begin{eqnarray} \label{L-iteration}
\ll_{ {\rm spinor} }( B ) &=&   \sum_{l=1}^{\infty}  2^l\, \ll_{{\rm
scalar}}( 2^{1-l} B)  \, . \label{iterated-b}
\end{eqnarray}
This is a completely new relation between spinor and scalar effective Lagrangians.

\begin{figure}[tb]
\begin{center}
\includegraphics[width=0.8\textwidth]{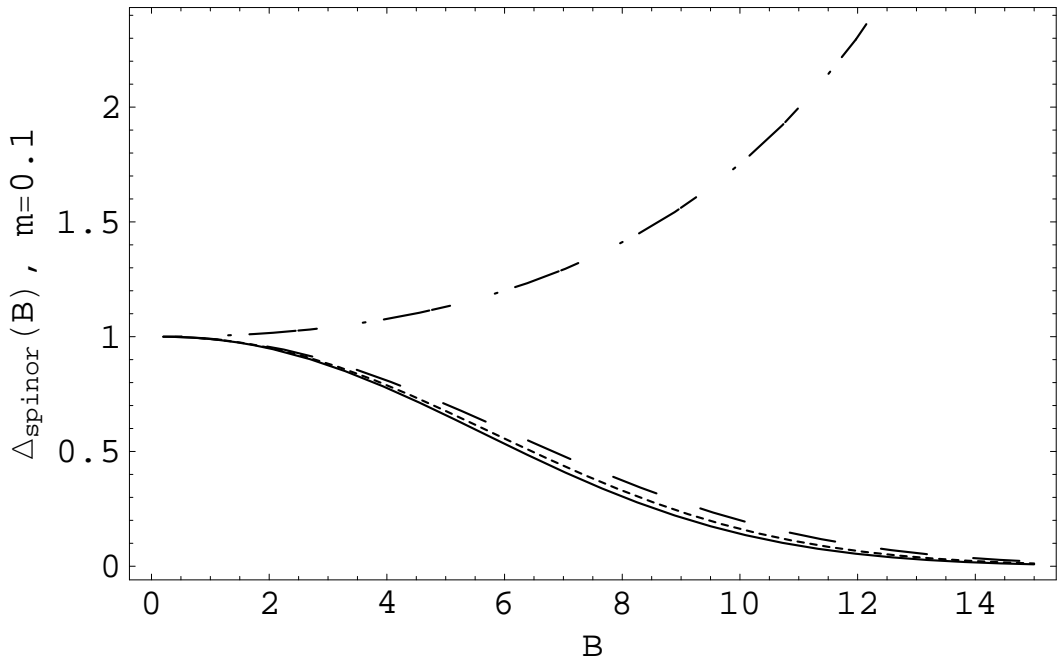}
\end{center}
\caption{The lowest curve (solid) corresponds to the exact spinor determinant,
 followed by  curves  resulting from only the first three terms (short-dashes) and the first two terms (long-dashes) of (\ref{det-iteration}).
Finally, the upper curve (dots-dashes), represents the exact scalar determinant. }
\label{fig1}
\end{figure}

This raises the question of convergence of  this series. Let us first reformulate the relation in terms
of functional determinants, we  have
\begin{eqnarray}
\Delta_{\rm spinor}\equiv \frac{\det\left[ \displaystyle{\not} \partial-e \displaystyle{\not} A + m\right]}{\det\left[ \displaystyle{\not} \partial+m\right]}
 = e^{-\int (\ll_{ {\rm spinor} }( B ) - \ll_{ {\rm spinor} }( 0 ) ) d^4 x}   \, ,
 \end{eqnarray}
 and
 \begin{eqnarray}
\Delta_{\rm scalar}\equiv \frac{\det\left[ \left(\partial_\mu -e A_\mu \right)^2+m^2\right]}{\det\left[\partial_\mu^2 + m^2\right]}
= e^{+\int (\ll_{ {\rm scalar} }( B ) - \ll_{ {\rm scalar} }( 0 ) ) d^4 x}   \, .
\end{eqnarray}
We can thus write (\ref{iterated-b}) as
\begin{eqnarray} \label{det-iteration}
\Delta_{\rm spinor}(B) &=& \prod_{l=1}^\infty (\Delta_{\rm scalar}
(2^{1-l} B) )^{- 2^l} \, .
\end{eqnarray}
This relation is illustrated in Figure \ref{fig1}, where we approximate the spinor determinant in terms of a finite product of scalar determinants. Note that just three terms already provide a  good approximation. The lower  curve shows the exact spinor determinant (solid) while the upper  curve (dot-dashes) is  the scalar determinant.

Note also  that Schwinger \cite{Schwinger:1954zz} has proven for general
electromagnetic background fields that the spinor determinant is bounded
above as $|\Delta_{\rm spinor}| \leq 1 $, whereas in the scalar case,
it has been argued \cite{Vafa:1983tf,Schrader:1978ny} that $|\Delta_{\rm scalar}| \geq 1$ . Both bounds
are consistent with our result~(\ref{det-iteration}), and the plots in Fig. \ref{fig1}.

It is also interesting to observe how this spinor/scalar relation manifests itself in the
context of one-loop  $\beta$-\emph{functions}.
As $B \to \infty$, by definition,  we have~\cite{Dunne:2002ta}
\begin{eqnarray}
\ll_{ {\rm spinor} }(B)  &\sim&  \beta_{{\rm spinor}} \frac{B^2}{2} \ln B \, , \\
\ll_{ {\rm scalar} }(B) &\sim&  \beta_{{\rm scalar}} \frac{B^2}{2} \ln B \, \, ,
\end{eqnarray}
where $\beta_{{\rm spinor}}$ and  $\beta_{{\rm scalar}}$ are the coefficients of the one-loop $\beta$-function
for spinor and scalar QED, respectively.
In the case of a constant magnetic background this gives $\beta_{{\rm spinor}} = \frac{1}{12 \pi^2}$
and $\beta_{{\rm scalar}} \frac{1}{48 \pi^2}$.
The strong-field limit of the spinor/scalar relation is easily obtained  from (\ref{L-iteration}), yielding
\begin{equation}
\beta_{{\rm spinor}} \frac{B^2}{2} \ln B \sim \beta_{{\rm scalar}}
\sum_{l = 1}^{\infty}  2^l    \frac{(2^{1- l} B)^2}{2} \ln (2^{1-l}
B) \, .
\end{equation}
The leading large $B$ behavior implies the correct relation between the one-loop $\beta$-function coefficients~:
\begin{equation}
\beta_{{\rm scalar}} = \frac{1}{4}  \beta_{{\rm spinor}} \, .
\end{equation}

\subsection{The case of a purely electric background}
\label{electric} The expressions for the spinor and scalar effective
Lagrangians under a constant electric field are
\begin{eqnarray}
\ll_{{\rm spinor}}(E) &=& - \frac{ E}{8\pi^2}\, \int_{0}^\infty \,
\frac{ds}{s^2} \; {\rm e}^{-m^2 s} \; \left( {\rm cot}\,   E s   -
\frac{1}{ E s} - \frac{ E s}{3} \right)\, ,
\label{2lsp}\\
\ll_{{\rm scalar}}(E)
&=&
 \frac{E}{16\pi^2}\, \int_{0}^\infty \, \frac{ds}{s^2}
\; {\rm e}^{-m^2 s} \; \left( \frac{1}{{\rm sin}\, E s}   -
\frac{1}{E s} + \frac{E s}{6} \right) \, . \label{2lsc}
\end{eqnarray}
As we see, the  electric field case is obtained from the constant magnetic
field case by the replacement $B\to i\, E$, so the same argument leads to the same  relation~:
\begin{eqnarray}
\ll_{{\rm spinor}}(E) &=& \frac{1}{2} \, \ll_{{\rm spinor}}(2 E) -
\ll_{{\rm scalar}}(2 E)  \, . \label{basic-e}
\end{eqnarray}
Iterating this basic relation (\ref{basic-e}), we obtain, as before
\begin{eqnarray}
\ll_{ {\rm spinor} }( E ) &=&   \sum_{l=1}^{\infty}  2^l\, \ll_{{\rm
scalar}}( 2^{1-l} E) \, .
\label{iterated-e}
\end{eqnarray}
Physically, however, there is a big difference between the electric and magnetic cases, as the electric background leads to pair production from vacuum, which is encoded in the imaginary part of $\ll$. The pair production rate is obtained after integrating
over the proper-time, and gathering the contributions from the poles of ${\rm cot}\, (E s)$  and $1/{\rm sin}\, (E s)$~:
 \begin{eqnarray}
{\rm Im}\, \ll_{{\rm spinor}}(E) &=& \frac{ E^2}{8 \pi^3} \sum_{n = 1}^\infty
\frac{1}{n^2} \, {\rm exp}\left[  \frac{- m^2 \pi n}{E} \right]
\label{imag-sp}
\end{eqnarray}
\begin{eqnarray}
{\rm Im}\, \ll_{{\rm scalar}}(E)
&=&  \frac{ E^2}{16 \pi^3} \sum_{n = 1}^\infty
\frac{(-1)^{n-1} }{n^2} \, {\rm exp}\left[  \frac{- m^2 \pi n}{E} \right]
\label{imag-sc}
\end{eqnarray}
Thus it is easy to see that~:
\begin{eqnarray}
\frac{1}{2} \; {\rm Im}\, \ll_{{\rm spinor}}(2 E) - {\rm Im}\, \ll_{{\rm scalar}}(2 E)  &=&
\frac{ (2 E)^2}{16 \pi^3} \sum_{n = 1}^\infty
\frac{1 - (-1)^{n-1} }{n^2}\,
{\rm exp}\left[  \frac{- m^2 \pi n}{(2E)} \right]    \nn \\
&=&
 \frac{ (2 E)^2}{8 \pi^3} \sum_{n = 2,\,4,\,6,\cdots}^\infty
\frac{1 }{n^2} \, {\rm exp} \; \left[  \frac{- m^2 \pi n}{ (2E)} \right] \nn   \\
&=&  \frac{ E^2}{8 \pi^3} \sum_{n = 1}^\infty
\frac{1 }{ n^2}\,  {\rm exp} \left[  \frac{- m^2 \pi n}{E} \right] \nn  \\
&=& {\rm Im}\, \ll_{{\rm spinor}}( E)
\label{basic-imag}
\end{eqnarray}
which is consistent with (\ref{basic-e}).
The alternating sign in the imaginary part of the scalar action has the
effect of cancelling the odd terms of the spinor action and the
resulting expression  is identifiable as another spinor Lagrangian.

\subsection{The case of a general constant electromagnetic
background} \label{electromagnetic}

For a general constant field,   the spinor and scalar Lagrangians are
given by (\ref{EHsp}) and (\ref{EHsc}).
 We find the basic doubling relation:
\begin{eqnarray}
\ll_{{\rm scalar}}(2 E, 2B)
&=& \ll_{{\rm spinor}}( 2 E, B) - \frac{1}{2} \ll_{{\rm spinor}}(2 E, 2 B)  \nn\\
&-& 2 \ll_{{\rm spinor}}( E , B)
+ \ll_{{\rm spinor}}( E , 2 B) \, ,
\label{basic-eb}
\end{eqnarray}
which follows immediately from the trigonometric identity
\begin{eqnarray}
\frac{1}{{\rm sin}(x){\rm sinh}(y)} &=&
{\rm cot}(x){\rm coth}(y) -  {\rm cot}(x/2){\rm coth}(y) \nn  \\
&+& {\rm cot}(x/2) {\rm coth}(y/2) - {\rm cot}(x){\rm coth}(y/2)  \, ,
\label{trig-2}
\end{eqnarray}

Note that in addition to the trigonometric functions appearing in this identity, each of the Lagrangians (\ref{EHsp}) and (\ref{EHsc}) contains renormalization terms. Remarkably, when we
write both sides  of (\ref{basic-eb}), these terms
coincide with (\ref{trig-2}),  as required.

An interesting consequence of (\ref{basic-eb}) is that it expresses the SUSY combination purely in terms of spinor Lagrangians~:
\begin{eqnarray}
\ll_{\rm N=2\; SUSY}(E, B)
&\equiv& \ll_{{\rm spinor}}( E, B) + 2 \ll_{{\rm scalar}}( E, B) \nn \\
&=&2 \ll_{{\rm spinor}}( E, B/2) -4 \ll_{{\rm spinor}}( E/2 , B/2)
+2 \ll_{{\rm spinor}}( E/2 , B) \nn \, .
\label{susy-eb}
\end{eqnarray}
%
%
%
%
%
%
%
%
%
%
%
%
%
%
%
%
%

\section{Zeta Function and  Gamma Function representations}
\label{multiplegamma}

Using different representations of the effective action is a way to reveal further spinor-scalar relations.
Other methods for computing functional determinants involve zeta functions. Given an operator $\mathcal{O}$, its
$\zeta$-function is defined as
\begin{equation}
\zeta (s) = {\rm tr}(\mathcal{O}^{-s})= \sum_{\lambda} \lambda^{-s}
\, .
\end{equation}
This allows one to write the functional determinant as :
\begin{equation}
{\rm {det}} \mathcal{O} = {\rm {exp}}[- \zeta^{'} (0)]
\end{equation}
The spinor Euler-Heisenberg effective action in a constant magnetic background may be written in terms of
the Hurwitz $\zeta$-function defined as
\begin{equation}
\zeta_H (s, z) = \sum_{n=0}^\infty (n + z )^{-s} \,.
\end{equation}
Like the Riemann zeta function $\zeta_{R}(s) \equiv \zeta_H (s, 1)$,
the Hurwitz  zeta function has an analytic continuation throughout
the entire complex $s$ plane with only a pole at $s=1$.
 In the case of a magnetic background the eigenvalues of the Dirac operator are
\begin{equation}
\lambda_n^{\pm} = m^2 + k_{\bot}^2 + e B (2 n + 1 \pm 1) \quad , \quad  n=0,1,\cdots
\end{equation}
where the  $\pm$ refers to different spin components and $\vec{k}_{\bot}$ is the transverse momentum.
The corresponding $\zeta$-function is
\begin{eqnarray*}
\zeta_{{\rm {spinor}}} (s) &=& \frac{e B}{2 \pi} \sum_{n = 0}^{\infty} \sum_{\pm} \int \frac{d^2 k_{\bot} }{(2 \pi)^2}
\Bigg( \frac{m^2 + k_{\bot}^2 + e B (2 n + 1 \pm 1)}{\mu^2}   \Bigg)^{-s} \nn \\
&=& \frac{m^4}{4 \pi^2} \Bigg( \frac{e B}{m^2} \Bigg)^2 \frac{(\frac{\mu^2}{2 e B})^s}{s - 1}
\Bigg[ 2 \zeta_H \Bigg(s-1, \,\frac{m^2}{2 e B}  \Bigg)  -  \Bigg( \frac{m^2}{2 e B} \Bigg)^{1-s}  \Bigg]
\end{eqnarray*}
where the scale $\mu$ is introduced to make the eigenvalues dimensionless. Using on-shell renormalization $(\mu = m)$ and
subtracting the zero-field contribution
 the one-loop effective Lagrangian is obtained~:
\begin{equation} \label{hurwitzsp}
\ll_{\rm {spinor}} =  \frac{e B}{2 \pi^2} \Bigg\{  \zeta_H^{'} \Bigg(-1, \,\frac{m^2}{2 e B}  \Bigg)
+  \zeta_H \Bigg(-1, \,\frac{m^2}{2 e B}  \Bigg)\ln \Bigg(\frac{m^2}{2 e B}  \Bigg)
-\frac{1}{12} + \frac{1}{4} \Bigg(\frac{m^2}{2 e B}  \Bigg)^2 \Bigg\} \, .
\end{equation}
We find useful to express the one-loop effective Lagrangian in terms of the
gamma function~\cite{Dunne:2004nc}.
The Hurwitz $\zeta$-function and the logarithm of the gamma function
are related in the following way~:
\begin{equation}
\zeta_H^{'} (-1, z) = \zeta^{'} (-1) - \frac{z}{2} \ln (2 \pi) - \frac{z}{2} (1-z) + \int_0^{z} \ln \Gamma(x) dx \, ,
\end{equation}
substituting this in (\ref{hurwitzsp}) yields
\begin{multline} \label{gammasp}
\ll_{\rm {spinor}} =  \frac{(e B)^2}{2 \pi^2} \Bigg\{   -\frac{1}{12} + \zeta^{'} (-1)
- \frac{m^2}{4 e B} + \frac{3}{4} \Bigg( \frac{m^2}{2 e B}  \Bigg)^2 - \frac{m^2}{4 e B} \ln (2 \pi) \\
+  \Bigg[ -\frac{1}{12}  + \frac{m^2}{4 e B} + \frac{1}{2} \Bigg( \frac{m^2}{2 e B}  \Bigg)^2   \Bigg]
\ln \Bigg( \frac{m^2}{2 e B}  \Bigg) + \int_0^{\frac{m^2}{2 e B}} \ln \Gamma(x) dx
 \Bigg\} \, .
\end{multline}
A similar expression holds for scalar QED
\begin{multline} \label{gammasc}
\ll_{\rm {scalar}} =  -\frac{(e B)^2}{4 \pi^2} \Bigg\{   \frac{5}{4} \Bigg( \frac{m^2}{2 e B}  \Bigg)^2
+ \Bigg[ \frac{1}{24} -  \frac{1}{2} \Bigg( \frac{m^2}{2 e B}  \Bigg)^2    \Bigg]
 \Bigg[ 1 + \ln  \Bigg( \frac{m^2}{2 e B}  \Bigg)   \Bigg] \\
-\frac{1}{2} \zeta^{'} (-1) - \frac{\ln 2}{24}  + \int_0^{\frac{m^2}{2 e B}} \ln \Gamma(x + \frac{1}{2}) dx
 \Bigg\} \, .
\end{multline}
We can derive the relation (\ref{basicB}) from the properties of the gamma function.
Note that
\begin{equation}
\ll_{\rm {spinor}} =  \frac{(e B)^2}{2 \pi^2}  \int_0^{\frac{m^2}{2 e B}} \ln \Gamma(x) dx + ({\rm {other \, terms}}) \, ,
\end{equation}
 while
 \begin{equation}
\ll_{\rm {scalar}} =  -\frac{(e B)^2}{4 \pi^2}  \int_0^{\frac{m^2}{2 e B}} \ln \Gamma(x + 1/2) dx + ({\rm {other \, terms}}) \, .
\end{equation}
In this form, the basic doubling relation (\ref{basicB}) results from the following \emph{duplication formula}
\begin{equation}
\Gamma (x) \Gamma (x + 1/2) = 2^{\frac{1}{2} - 2 x} e^{- \zeta^{'} (0)} \Gamma (2x) \, .
\end{equation}
The spinor/scalar  relation (\ref{iterated-b}) is derived by
iteration, as before.
\section{Spinor/Scalar relations in Curved Spacetime}
\label{gravity} So far we have shown how one can find relations
between the spinor and scalar effective Lagrangians using different
representations. In particular we have used the representation of
the effective action in terms of $\ln \Gamma (x)$. The \emph{digamma
function} is defined as
\begin{equation}
\psi (x) \equiv \frac{d}{dx} \ln \Gamma(x) \, .
\end{equation}
This type of function is common in the effective actions of
different gauge-theories.

We now turn our attention to the effective Lagrangians of  spinor
and scalar particles in a curved two-dimensional AdS space.  In this
scenario the  Dirac and Klein-Gordon operators get modified by gauge
terms that account for the curvature of space. Let us write the
spinor and scalar effecive Lagrangians for a particle in a
two-dimensional AdS following  \cite{Camporesi:1992tm} and
 \cite{Kamela:1998mb}.

For a spin-$\frac{1}{2}$ particle  the coincident propagator is
obtained from the trace of the inverse Dirac operator
\begin{equation}
\frac{\partial \ll}{\partial m^2} = -{\rm {Tr}} \Bigg( \frac{1}{
\displaystyle{\not} \nabla - m} \Bigg)_{{\rm AdS}} = \frac{
-\lambda^{\frac{d-1}{2}} \, 2^{\frac{d}{2}} \, \Gamma (1 -
\frac{d}{2})  \Gamma (\frac{d}{2} + \sqrt{m^2 / \lambda })   }{  (4
\pi)^{\frac{d}{2}}  \, \Gamma (1 - \frac{d}{2} + \sqrt{m^2 / \lambda
})     } \, ,
\end{equation}
where  $\lambda$ is given by the Ricci scalar and represents the curvature of space and $d$
is the number of space-time dimensions.
Expanding around $d=2$ we get
\begin{equation}
-{\rm {Tr}} \Bigg( \frac{1}{ \displaystyle{\not} \nabla - m}
\Bigg)_{{\rm AdS}_2} = \frac{m}{\pi (d-2)} + \frac{m}{\pi} \Bigg[
\psi \Bigg( \sqrt{\frac{m^2}{\lambda}} \Bigg)
 + \frac{1}{2} \ln \Bigg( \frac{\lambda}{2 \pi}   \Bigg) + \frac{\gamma}{2} \Bigg] + \frac{\sqrt{\lambda}}{2 \pi}
+ \mathcal{O}(d-2) \, . \nn
\end{equation}
Thus the  effective Lagrangian is
\begin{equation}
\ll_{ {\rm {AdS}}_2  }^{\rm {spinor}} (\sqrt{m^2/\lambda}) =
\frac{1}{ \pi} \int dm \, m \, \psi ( \sqrt{m^2/ \lambda}) \, ,
\end{equation}
after rescaling we get
\begin{equation} \label{d2AdS}
\ll_{ {\rm {AdS}}_2  }^{\rm {spinor}} (\sqrt{m^2/\lambda}) =
\frac{\lambda}{ \pi} \int_{0}^{\sqrt{m^2/ \lambda}} dy \, y \, \psi
(y) \, .
\end{equation}
 In the case of the
scalar effective Lagrangian we have the following operator
\begin{equation}
\mathcal{H} = -\Box_g + m^2 \, ,
\end{equation}
where $\Box_g = \frac{1}{\sqrt{-g}} \partial_\mu (g^{\mu
\nu}\sqrt{-g} \partial_\nu)$. In the case $d=2$ the coincident
propagator is
\begin{equation}
\frac{i}{2} \mathcal{G}(x,x) = \frac{1}{4 \pi (n-2)}- \frac{1}{8
\pi} \Bigg[\ln \Bigg( \frac{4 \pi \Lambda}{\lambda} \Bigg) - \gamma
- 2 \, \psi   \Bigg(  \frac{1}{2} + \frac{1}{2} \sqrt{1 + \frac{4
m^2}{\lambda}} \Bigg) \Bigg]
\end{equation}
and the corresponding effective Lagrangian is
\begin{equation}
\ll_{ {\rm {AdS}}_2  }^{\rm {scalar}} (\sqrt{m^2/\lambda}) =
\frac{1}{4 \pi} \int d(m^2) \; \psi   \Bigg(  \frac{1}{2} +
\frac{1}{2} \sqrt{1 + \frac{4 m^2}{\lambda}} \Bigg)
\end{equation}
or after changing the variable
\begin{equation} \label{lgravsc}
\ll_{ {\rm {AdS}}_2  }^{\rm {scalar}}(\sqrt{m^2/\lambda}) =
\frac{\lambda}{4 \pi} \int_{1/2}^{\sqrt{\frac{m^2}{\lambda} +
\frac{1}{4} }} dy \; y \, \psi \Bigg( y + \frac{1}{2} \Bigg) \, .
\end{equation}
We use the following identity to connect the spinor and scalar
effective actions.
\begin{equation}
\psi (x + 1/2) + \psi (x) - 2 \psi (2 x) + 2 \ln 2 = 0
\end{equation}
Substituting this equation into (\ref{lgravsc}) we get
\begin{eqnarray}
\ll_{ {\rm {AdS}}_2  }^{\rm {scalar}}(\sqrt{m^2/\lambda}) &=& -
\frac{ \lambda}{4 \pi } \int_{1/2}^{\sqrt{\frac{m^2}{\lambda} +
\frac{1}{4} }} dy \; y \, \psi (y)
+  \frac{\lambda}{4 \pi} \int_{1/2}^{\sqrt{\frac{m^2}{\lambda} +
\frac{1}{4} }} dy \; (2 y) \, \psi (2 y) \nn \\
&-& \frac{\lambda (2 \ln 2)}{\pi} \int_{1/2}^{\sqrt{\frac{m^2}{\lambda}
+ \frac{1}{4} }}\; y \, dy \, .
\end{eqnarray}
Rescaling the integration variable in the first two terms we write
the right-hand side in terms of spinor Lagrangians :
\begin{eqnarray}
\ll_{ {\rm {AdS}}_2  }^{\rm {scalar}}(\sqrt{m^2/\lambda}) &=& -
\frac{1}{4}\cdot \frac{\lambda}{\lambda^{'}} \ll_{ {\rm {AdS}}_2
}^{\rm {spinor}}(\sqrt{m^2/\lambda^{'}}) + \frac{1}{8} \cdot
\frac{\lambda}{\lambda^{''}}\ll_{ {\rm {AdS}}_2 }^{\rm
{spinor}}(\sqrt{m^2/\lambda^{''}}) - \Bigg( \frac{\ln 2}{4
\pi}\Bigg) m^2 \, , \nn \\
\end{eqnarray}
where
\begin{equation}
\lambda^{'} = \frac{\lambda}{1 + \frac{\lambda}{4 m^2}}
\end{equation}
and
\begin{eqnarray}
\lambda^{''} &=& \frac{1}{4} \Bigg( \frac{\lambda}{1 + \frac{\lambda} {4 m^2}} \Bigg) \, .
\end{eqnarray}
We  write the last expression as
\begin{eqnarray}
\ll_{ {\rm {AdS}}_2  }^{\rm {scalar}}(\sqrt{m^2/\lambda}) &=& -
\frac{1}{4}\cdot \frac{\lambda}{\lambda^{'}} \ll_{ {\rm {AdS}}_2
}^{\rm {spinor}}(\sqrt{m^2/\lambda^{'}}) + \frac{1}{2} \cdot
\frac{\lambda}{\lambda^{'}}\ll_{ {\rm {AdS}}_2 }^{\rm
{spinor}}(2 \sqrt{m^2/\lambda^{'}}) - \Bigg( \frac{\ln 2}{4
\pi}\Bigg) m^2 \, , \nn \\
\end{eqnarray}
Since $\frac{\lambda^{'}}{\lambda} = 1 - \frac{\lambda^{'}}{ 4 m^2}$ we can write the basic relation as
 \begin{eqnarray}
\ll_{ {\rm {AdS}}_2  }^{\rm {spinor}}(x) &=& 2
 \, \ll_{ {\rm {AdS}}_2
}^{\rm {spinor}}(2 x) - 4 \Bigg( 1 - \frac{1}{ 4 x^2} \Bigg) \ll_{ {\rm {AdS}}_2 }^{\rm
{scalar}}\Bigg( x \Big( 1 - \frac{1}{ 4 x^2} \Big)^{\frac{1}{2}} \Bigg)  \, ,
\end{eqnarray}
where $x \equiv \sqrt{m^2/\lambda^{'}}  $ and we have omitted the mass term.

Iterating this relation $N$ times we obtain
\begin{eqnarray}
\ll_{ {\rm {AdS}}_2  }^{\rm {spinor}}(x) &=& 2^{N + 1} \ll_{ {\rm {AdS}}_2
}^{\rm {spinor}}(2^{N + 1} x) - \sum_{n = 0}^N 2^n  4 \Bigg( 1 - \frac{1}{ 4 (2^n x)^2} \Bigg)
\ll_{ {\rm {AdS}}_2 }^{\rm
{scalar}}\Bigg( (2^n x) \Big( 1 - \frac{1}{ 4 (2^n x)^2} \Big)^{\frac{1}{2}} \Bigg)  \nn \\
&& \nn \\
&&
\end{eqnarray}
Rescaling $x$ by $2^{-(N+1)}$, we obtain
\begin{equation}
\ll_{ {\rm {AdS}}_2  }^{\rm {spinor}}(x) = 2 \sum_{l = 0}^N 2^{-l} \Bigg( 1 - \frac{1}{2^{-2 l} x^2 } \Bigg) \ll_{ {\rm {AdS}}_2 }^{\rm
{scalar}} \Bigg(  2^{- (l + 1) }  x   \Big( 1 - \frac{1}{2^{-2 l} x^2 } \Big)^{ \frac{1}{2} }  \Bigg) \, .
\end{equation}

\section{Conclusion}

In this paper I have presented simple new relations between the spinor and scalar one-loop
effective Lagrangians in both electromagnetic and gravitational backgrounds. Since the effective action is the
generating function of scattering amplitudes~\cite{Martin:2003gb}, these relations may be used to relate the low energy limits of such
scattering amplitudes.


\end{document}